\documentclass[aps,pre,twocolumn,amsmath,amssymb,superscriptaddress,showpacs,longbibliography]{revtex4-1}

\usepackage{graphicx}
\usepackage{dcolumn}
\usepackage{bm}

\usepackage{natbib}
\usepackage[
text={7in,9.5in},centering,
]{geometry}

\usepackage{epsfig}
\usepackage{amsmath}
\usepackage{amssymb}
\usepackage{textcomp}

\begin{document}

\title{Scale-free networks with exponent one}

\author{G. Tim\'ar}
 \email{gtimar@ua.pt}
 \affiliation{Departamento de F\'\i sica da Universidade de Aveiro \& I3N, Campus Universit\'ario de Santiago, 3810-193 Aveiro, Portugal}

\author{S. N. Dorogovtsev}
 \affiliation{Departamento de F\'\i sica da Universidade de Aveiro \& I3N, Campus Universit\'ario de Santiago, 3810-193 Aveiro, Portugal}
 \affiliation{A. F. Ioffe Physico-Technical Institute, 194021 St. Petersburg, Russia}

\author{J. F. F. Mendes}
 \affiliation{Departamento de F\'\i sica da Universidade de Aveiro \& I3N, Campus Universit\'ario de Santiago, 3810-193 Aveiro, Portugal}

\date{\today}

\begin{abstract}
A majority of studied models for scale-free networks have degree distributions with exponents greater than $2$. Real networks, however, can demonstrate essentially more heavy-tailed degree distributions. 
We 
explore two 
models of scale-free equilibrium networks that have the degree distribution exponent $\gamma = 1$, $P(q) \sim q^{-\gamma}$. 
Such degree distributions can be identified in empirical data only if the mean degree of a network is sufficiently high. 
Our models 
exploit a rewiring mechanism. 
They 
are local in the sense that no knowledge of the network structure, apart from the immediate neighbourhood of the vertices,
is required. 
These models generate uncorrelated networks in the infinite size limit, where they are solved explicitly. 
We investigate finite size effects by the use of simulations. 
We find that both models exhibit disassortative degree--degree correlations for finite network sizes. 
In addition, we observe a markedly degree-dependent clustering in the finite networks. 
We indicate a real-world network with a similar degree distribution.
\end{abstract}



\maketitle

\section{Introduction}

Scale-free networks have been in the forefront of networks research for almost two decades.  
Examples range from social networks \cite{newman2001scientific,
liljeros2001web}, biological networks \cite{albert2005scale, eguiluz2005scale} to artificial networks like the internet
\cite{faloutsos1999power} and the world-wide web \cite{albert1999internet, barabasi2000scale}.
The 
widely accepted mechanism for the evolution of growing scale-free networks is 
preferential attachment 
\cite{barabasi1999emergence}. 
This mechanism has been extensively 
explored 
and many generalizations and modifications of the original model have been suggested \cite{dorogovtsev2000structure,krapivsky2000connectivity,albert2002statistical,dorogovtsev2002evolution,dorogovtsev2003evolution}. The same principle has also been applied
to equilibrium networks \cite{dorogovtsev2003principles}.
The preferential attachment mechanism has 
relations with random multiplicative processes (\cite{adamic2000power,levy1996power,
sornette1998multiplicative, mitzenmacher2004brief}) which have relevance
in many fields of statistical physics, and are known to produce 
skewed distributions.
Besides preferential attachment, some other methods of producing scale-free networks have been exploited, for example,
fitness-based models \cite{bianconi2001competition,bianconi2001bose,caldarelli2002scale}, merging processes \cite{kim2005self,alava2005complex,seyed2006scale}, optimization models
\cite{valverde2002scale, d2007power}, urn-based statistical ensembles \cite{burda2001statistical,burda2003uncorrelated}, networks embedded into metric spaces \cite{krioukov2010hyperbolic,papadopoulos2012popularity}, and others.  

Most well-studied real networks appear to have degree distribution exponents larger than two. As a result, and also due to the
popularity of the preferential attachment mechanism, networks with smaller exponents have received much less attention.
In such networks the mean degree diverges, and consequently the ``natural'' cutoff of the degree distribution scales with the system
size in a different way compared to networks with higher exponents \cite{seyed2006scale}. The case of $\gamma = 1$ is even more
peculiar: the normalization condition implies that the cutoff of the degree distribution must remain finite in an infinite
system. This circumstance makes it rather difficult to clearly identify such distributions in empirical data obtained from sparse networks. On the other hand, if the mean vertex degree is sufficiently large, this kind of distribution can be observed (see below), which justifies our investigation. To our knowledge, the only studied model for networks with $\gamma = 1$, was that of trees growing based on ``the power of choice'' (local optimization of the new connections) \cite{d2007power}.
Note that variations of the models with hidden variables \cite{zuev2016hamiltonian, colomer2012clustering, anand2014entropy} can also generate such degree distributions if one chooses an appropriate distribution of hidden variables.

In the present paper we consider two simple equilibrium network models producing $\gamma = 1$. These rewiring models are local
in the sense that the vertices need to ``know'' only the structure of their immediate neighbourhoods. We show that the resulting
degree distribution has a simple exact solution in the sparse network limit, where the network is uncorrelated, which is a power-law of
$\gamma = 1$ with an exponential cutoff that is determined by the mean degree. We perform extensive numerical simulations of these models for finite networks, and observe marked disassortative degree--degree correlations.

\begin{figure}[t]
\centering
\includegraphics[width=200pt,angle=0.]{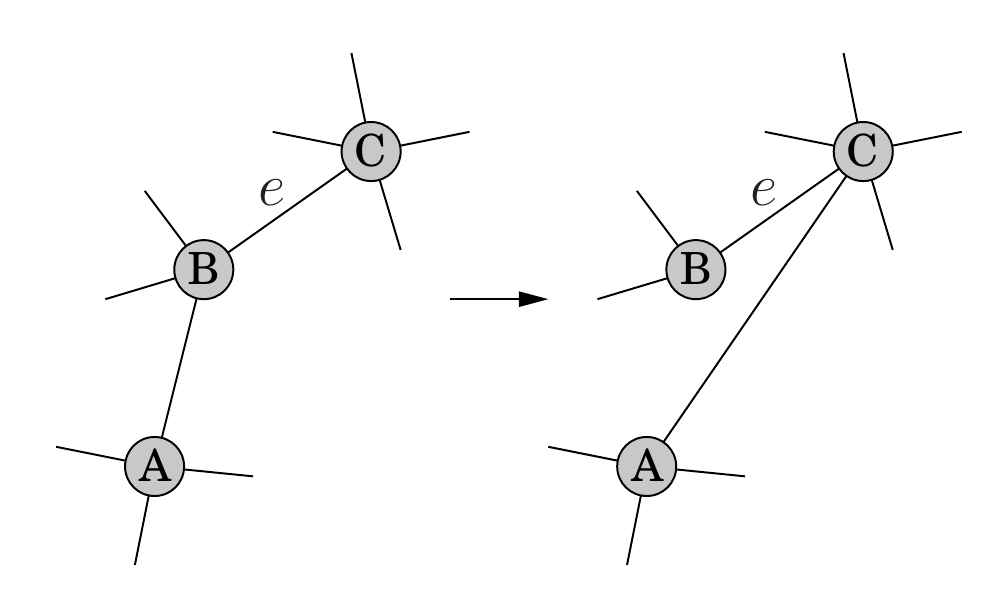}
\caption{Schematic representation of the evolution mechanism for models $1$ and $2$.}
\label{fig:diagram}
\end{figure}


\section{The models}

A well-studied class of mechanisms that is known to produce power-law distributions, is random multiplicative processes
\cite{adamic2000power,levy1996power, sornette1998multiplicative, mitzenmacher2004brief}.
The essence of such models is that the fluctuations of random variables 
are proportional to their values (independent fluctuations would result in classical Brownian dynamics). 
A simple example of such processes is discussed in
\cite{pietronero2001explaining} and is shown to generate power-law distributions of exponent 1.
We use a 
similar principle. In our case, the random variables are the degrees of nodes in the network. Instead of
fluctuations which are proportional to the values of the random variables, we apply fluctuations of a fixed size (rewiring one link
at a time) with a probability that is proportional to the values of the random variables (degrees of nodes).
Here we consider a model of equilibrium networks that 
realizes this scheme.

\textit{Model 1}. 
Consider an arbitrary connected graph. At every step of the evolution do the following:

\begin{enumerate}

 \item Choose an edge uniformly at random (edge $e$ in Fig.~\ref{fig:diagram}). 
 
 \item Reattach a neighbour (node $A$ in Fig. \ref{fig:diagram}) of one of its end nodes to the other
       (from node $B$ to node $C$ in Fig. \ref{fig:diagram}).
  
\end{enumerate}
Repeat the above procedure until equilibrium is reached.
In the second step node $A$ is chosen uniformly at random from the set of all neighbours of $B$ which are not neighbours of $C$
(and are not themselves $C$). If there are no such nodes, then nothing should be done in 
this iteration.

We denote the degree distribution by $P(q)$. The joint degree distribution, i.e., the probability that the end nodes of a uniformly randomly chosen link have degrees $q$ and $q'$ is denoted by $P(q,q')$, and the conditional probability that an end node of a random link has degree $q$ given that the other end node has degree $q'$, by $P(q|q')$.
In a given step of the evolution the probability of a node of degree $q$ to be chosen as node $B$ is 
\begin{equation}
   P_B(q) = \frac{1}{2} \frac{qP(q)}{\langle q \rangle}
  \label{eq:Pb0}
\end{equation}
and, similarly, the probability of a node of degree $q$ to be chosen as node $C$ is
\begin{equation}
   P_C(q) = \frac{1}{2} \frac{qP(q)}{\langle q \rangle} = P_B(q)
   .
  \label{eq:Pc0}
\end{equation}
Assuming that in equilibrium, clustering is purely a result of degree--degree correlations,
we introduce $R(q,q')$, the probability that if a node of degree $q$ is chosen as $B$ and a node of degree $q'$ as $C$,
then a rewiring is possible, i.e., that $B$ has at least one neighbour that is not a neighbour of $C$:
\begin{widetext}
\begin{eqnarray}
R(1,q') &=& 0
,
 \nonumber
   \\[5pt]
R(q>1,q') &=& 1 - \left\lgroup \sum_k P(k|q) P(q'|k) \frac{(q'-1)(k-1)}{Nq'P(q')} \right\rgroup ^{q-1} 
      \nonumber
   \\[5pt]
&=& 
   1 - \left\lgroup \frac{\langle q \rangle ^2 (q'-1)}{Nqq'P(q)P(q')} \sum_k \frac{P(k,q)P(q',k)(k-1)}{kP(k)} \right
  \rgroup ^{q-1} 
   , 
  \label{eq:R0}
\end{eqnarray}
\end{widetext}
where $N$ is the number of nodes in this network. 
Now we can write the probability that a node of degree $q$ is chosen as $B$ and rewiring is possible:
\begin{eqnarray}   
     P_{B,r}(1) &=& 0 
     ,
\nonumber
\\[5pt]
P_{B,r}(q>1) &=& \frac{1}{2} \frac{qP(q)}{\langle q \rangle} \sum_{q'} P(q'|q) R(q,q') 
. 
  \label{eq:PrewB}
\end{eqnarray}
Similarly, the probability that a node of degree $q$ is chosen as $C$ and rewiring is possible:
\begin{equation}
   P_{C,r}(q) = \frac{1}{2} \frac{qP(q)}{\langle q \rangle} \sum_{q'>1} P(q'|q) R(q',q)
   .
  \label{eq:PrewC}
\end{equation}
It is easy to see from Eq. (\ref{eq:R0}) that in the limit $N \rightarrow \infty$, $R(q,q')=1$ for any $q > 1,q'$ and
$R(1,q') = 0$. In this case, Eq. (\ref{eq:PrewB}) reduces to 
\begin{eqnarray}   
P_{B,r}(1) &=& 0
,
\nonumber
\\[5pt]
P_{B,r}(q>1) &=& \frac{1}{2} \frac{qP(q)}{\langle q \rangle} 
,
  \label{eq:PrewBr}
\end{eqnarray}
and Eq. (\ref{eq:PrewC}) becomes 
\begin{equation}
   P_{C,r}(q) = \frac{1}{2} \frac{qP(q)}{\langle q \rangle} b(q)
   ,
  \label{eq:PrewCr}
\end{equation}
where $b(q) = 1 - P(1|q)$.
Noting that in the stationary state the probability of a node of degree $q+1$ losing an edge must match
the probability of a node of degree $q$ gaining an edge, we can write the 
stationary equation for the degree distribution:
\begin{equation}
   P_{C,r}(q) = P_{B,r}(q+1)
   .
  \label{eq:Msimple0}
\end{equation}
Substituting Eqs. (\ref{eq:PrewBr}) and (\ref{eq:PrewCr}) into Eq. (\ref{eq:Msimple0}), we have:
\begin{equation}
   \frac{1}{2} \frac{qP(q)}{\langle q \rangle}b(q) = \frac{1}{2} \frac{(q+1)P(q+1)}{\langle q \rangle}
   .
  \label{eq:Msimple1}
\end{equation}
We see that degree--degree correlations appear only in $b(q)$. If we assume that $b(q)$ is constant (this is a 
weaker
assumption than assuming that correlations are entirely absent), then $b(q) = c = 1 - 1P(1)/\langle q \rangle$.
This can be easily seen in the following way. For any network, regardless of degree-degree correlations, the degree
distribution of end nodes of links is $qP(q) / \langle q \rangle$ (the probability, that a randomly chosen end node of
a randomly chosen link has degree $q$). The probability that an end node of a randomly chosen link has degree 1, is
therefore $1P(1) / \langle q \rangle$. This probability can also be written as:
\begin{equation}
   \frac{1P(1)}{\langle q \rangle} = \sum_q P(1|q) \frac{qP(q)}{\langle q \rangle}
   .
  \label{eq:leaf1}
\end{equation}
Assuming that $P(1|q)$ is constant ($P(1|q) = h$), we have 
\begin{equation}
   \frac{1P(1)}{\langle q \rangle} = \sum_q P(1|q) \frac{qP(q)}{\langle q \rangle} = \sum_q h \frac{qP(q)}{\langle q \rangle} = h
   .
  \label{eq:leaf2}
\end{equation}
Therefore $b(q) = 1 - P(1|q) = 1 - 1P(1) / \langle q \rangle = c$. 
Then Eq. (\ref{eq:Msimple1}) is simply 
\begin{equation}
   qP(q)c = (q+1)P(q+1)
   .
  \label{eq:Msimple2}
\end{equation}
The solution of Eq. (\ref{eq:Msimple2}) is the following:
\begin{equation}
   P(q) = A q^{-1} e^{- q/q_{\text{cut}}}
   .
  \label{eq:solution}
\end{equation}
where $q_{\text{cut}} = -1 / \ln c$. The cutoff $q_{\text{cut}}$ may also be expressed in terms of the mean
degree $\langle q \rangle$. The constant $A$ in Eq. (\ref{eq:solution}), from the normalization
condition, is
\begin{equation}
   A = \frac{1}{\sum_{q=1}^{\infty} q^{-1} e^{- q/q_{{cut}}} } = -\,\frac{1}{\ln(1 - e^{- 1/q_{\text{cut}}})} \cong \frac{1}{\ln q_{\text{cut}}}
   ,
  \label{eq:const}
\end{equation}
and the mean degree is 
\begin{equation}
   \langle q \rangle = A \sum_{q=1}^{\infty} e^{- q/q_{\text{cut}}} = \frac{A}{1 - e^{- 1/q_{\text{cut}}}} \cong \frac{q_{\text{cut}}}{\ln q_{\text{cut}}}
   , 
  \label{eq:meandeg}
\end{equation}
so 
$q_{\text{cut}}$ 
\begin{equation}
q_{\text{cut}}   \cong  \langle q \rangle \ln \langle q \rangle 
  \label{160}
\end{equation}
is independent of the system size.

Using Eq. (\ref{eq:R0}), it is possible in principle  to write the complete master equation for the degree--degree distribution $P(q,q')$ in the general
case and solve it numerically. However, this undertaking would be very cumbersome, therefore, instead, in
Sec.~\ref{sec:sim} we present simulation results. 
Let us first 
introduce an alternative formulation of the above model.

\textit{Model 2.} 
Consider an arbitrary connected graph. At every step of the evolution do the following:

\begin{enumerate}

 \item Choose a node uniformly at random (node $A$ in Fig. \ref{fig:diagram}). 
 
 \item Choose a neighbour of $A$ uniformly at random (node $B$ in 
 this figure). 
 
 \item Choose a $2^{nd}$ neighbour of $A$ through $B$ (node $C$) 
       uniformly at random from all $2^{nd}$ neighbours of $A$ through $B$.
       If no such node exists, do nothing in this iteration.
 
 \item Rewire $A$ from $B$ to $C$  (as in the figure). 
 
\end{enumerate}
Repeat the above procedure until equilibrium is reached. 
In other words, a 
node rewires its connection from a randomly chosen nearest neighbour to a ``descendant'' of this neighbour.   
We can write the probability $P_B(q)$ that in a given step of the evolution a node of degree $q$ is selected to be node $B$: 
\begin{equation}
   P_B(q) = \sum_{q'} P_A(q') P(B{:}q | A{:}q') = \langle q \rangle \sum_{q'} \frac{P(q,q')}{q'}
   ,
  \label{eq:Pb2}
\end{equation}
where $P(B{:}q | A{:}q')$ is the conditional probability that a node of degree $q$ is chosen to be node $B$, given that a node of degree $q'$ was chosen to be node $A$.
Assuming, again, that in equilibrium, clustering is purely a result of degree-degree correlations, and considering the
limit $N \rightarrow \infty$ (i.e. assuming that clustering goes to zero), we can write the probability $P_{BC}(q,q')$ that
at a given step a node of degree $q$ is chosen as $B$ and a node of degree $q'$ as $C$:
\begin{eqnarray}
\!\!\!\!\!
P_{BC}(1,q')&=&0
,
   \nonumber
   \\[5pt]
\!\!\!\!\!   
P_{BC}(q>1,q')&=&P_B(q) P(C{:}q' | B{:}q) =
   \nonumber
   \\[5pt]
\!\!\!\!\!   
&=& \langle q \rangle \sum_{q''} \frac{P(q,q'')}{q''} \frac{P(q,q')}{qP(q)} \langle q \rangle =
   \nonumber
   \\[5pt]
\!\!\!\!\!
&=& \langle q \rangle ^2 \frac{P(q,q')}{qP(q)} \sum_{q''} \frac{P(q,q'')}{q''}
   .
  \label{eq:Pbc}
  \end{eqnarray}
Again, requiring that in the stationary state the probability of a node of degree $q+1$ losing an edge match the probability
of a node of degree $q$ gaining an edge, we can write the equation:
\begin{equation}
   \sum_{q'} P_{BC}(q+1,q') = \sum_{q'} P_{BC}(q',q)
   .
  \label{eq:Mcomplex}
\end{equation}
We see that even in the limit of infinite size, the situation is much more complex than Eq. (\ref{eq:Msimple1}). If we further assume
that if $N \rightarrow \infty$, then the equilibrium network is uncorrelated, i.e., that the degree-degree distribution factors,
we find that Eq. (\ref{eq:Pb2}) reduces to
\begin{equation}
   P_B(q) = \frac{qP(q)}{\langle q \rangle}
   . 
  \label{eq:Pb2_r}
\end{equation}
Then Eq. (\ref{eq:Pbc}) takes the simple form:
\begin{eqnarray}
P_{BC}(1,q') &=& 0
  ,
   \nonumber
 \\[5pt]
P_{BC}(q>1,q') &=& \frac{qP(q)q'P(q')}{\langle q \rangle ^2}
,
  \label{eq:Pbc_r}
  \end{eqnarray}
and the stationary  equation (Eq. (\ref{eq:Mcomplex})) is now simply 
\begin{equation}
   \frac{(q+1)P(q+1)}{\langle q \rangle} = \frac{qP(q)}{\langle q \rangle} c
  \label{eq:Msimple3}
\end{equation}
with $c = 1 - 1P(1)/\langle q \rangle$, which is identical to Eq. (\ref{eq:Msimple2}).

We see that in the limit 
$N \rightarrow \infty$, assuming that the network is then uncorrelated, the two model
formulations are equivalent.
Models 1 and 2 are closely related, exploiting the same mechanism.
While model 1 is in fact a null model, there is a rationale behind model 2.
In this model, a node redirects one of its connections to get the farthest possible reach by using only local information
from its nearest neighbours (the lists of their neighbours). The redirection of a link, instead of the addition of a new one,
corresponds to evolution with limited resources.
In the following we employ simulations to investigate the behaviour of the two models
in a wide range of system sizes. The simulations indicate that in the infinite sparse network limit, both formulations
lead to uncorrelated equilibrium networks, although the models are significantly different for finite networks.


\section{Simulations} \label{sec:sim}

We performed simulations of varying system sizes and mean degrees, averaging over at least 10 realizations for each combination
of parameters.
The starting graph in each simulation was a connected random graph generated in the following way: first
all the nodes were linked in a chain to ensure connectedness, then all remaining links were assigned to the nodes randomly.
$T=10^{10}$ time steps (rewiring attempts) were used in each simulation, this ensured that equilibrium was reached in each case.
The success rate for rewiring was above $95\%$ for all parameter settings.
First we investigate the case of $N \gg \langle q \rangle$, approaching the limit of large sparse networks.
Clustering and degree-degree correlations are found diminishing as this limit is approached. Secondly we analyse more dense
networks to compare the behaviour of the two model formulations.

\begin{figure}[t]
\centering
\includegraphics[width=200pt,angle=0.]{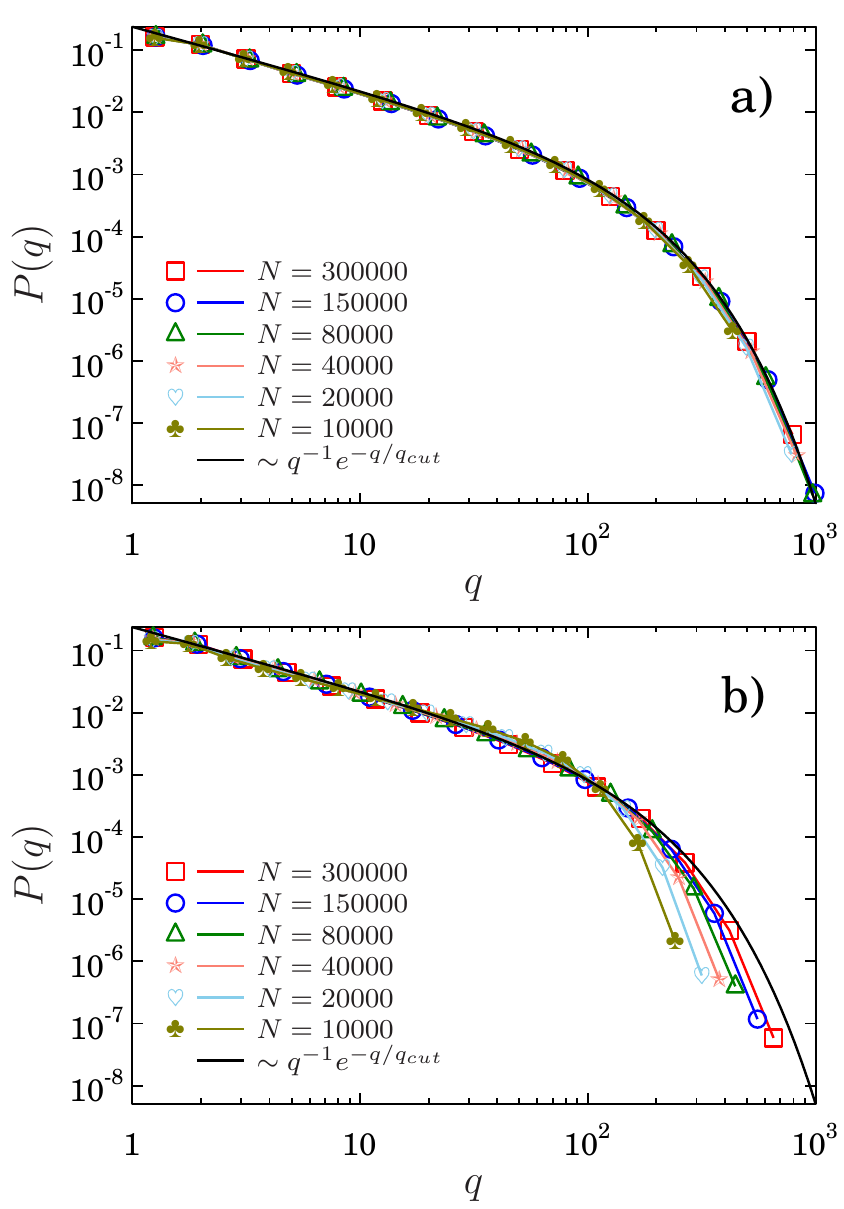}
\caption{(Color online) Degree distribution of sparse equilibrium networks of mean degree $\langle q \rangle = 20$ and different sizes for model $1$ (a) and model $2$ (b).
}
\label{fig:sparse_dd}
\end{figure}


\subsection{Sparse networks}

Degree distributions at equilibrium are shown in Fig.~\ref{fig:sparse_dd} 
for both models, for different system sizes, and fixed mean degree $\langle q \rangle = 20$.
In both cases, for large system sizes, the uncorrelated form of the degree distribution Eq.~(\ref{eq:solution}), is approached,
but this convergence is 
much slower for the second model. 
The choice of mean degree was limited by the system size for which simulations run in reasonable time for the computationally more demanding model 2.

\begin{figure}[t]
\centering
\includegraphics[width=200pt,angle=0.]{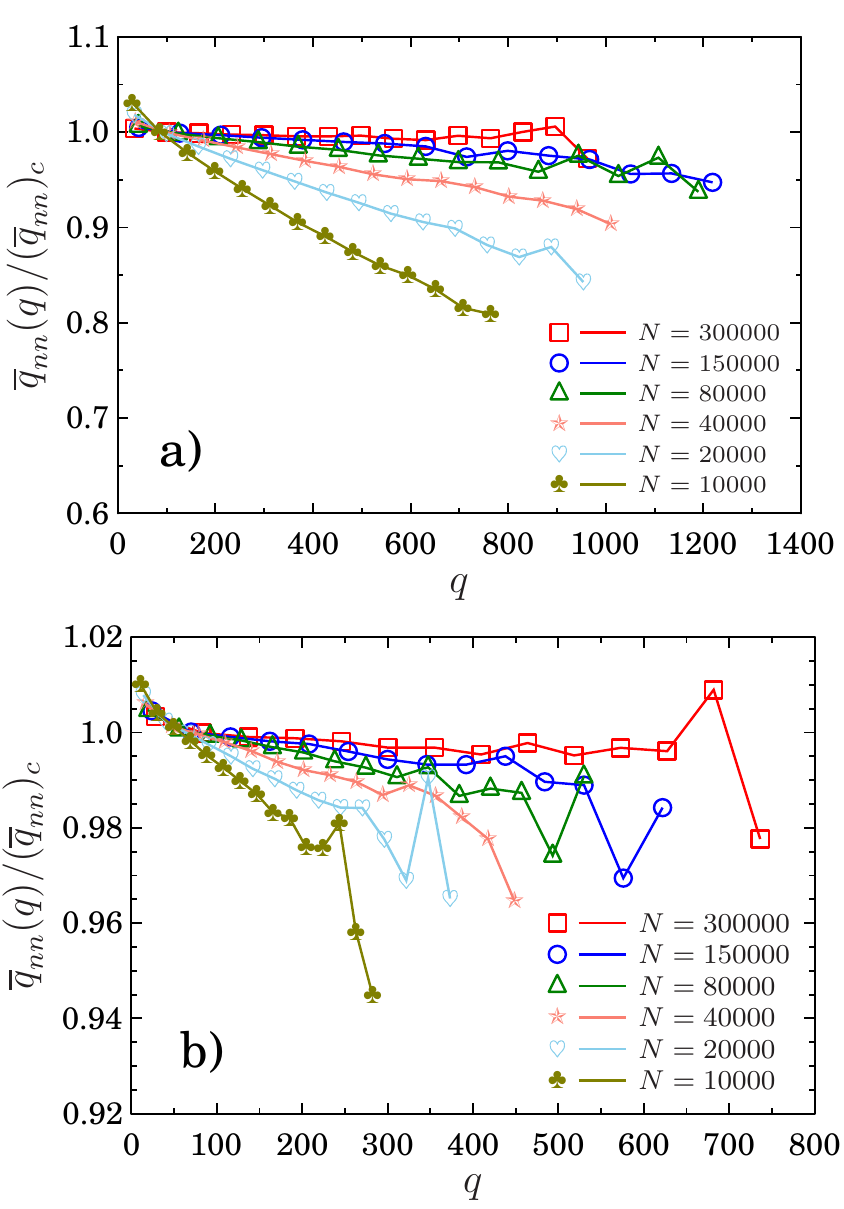}
\caption{(Color online) Degree dependence of the mean degree $\overline{q}_{nn}$ of the nearest neighbours of a node of degree $q$ for sparse equilibrium networks of mean degree $\langle q \rangle = 20$
and different sizes. (a) model 1, (b) model 2. The mean degree $\overline{q}_{nn}$ is normalized by its value for the corresponding uncorrelated network.}
\label{fig:sparse_qnn}
\end{figure}

\begin{figure}[!]
\centering
\includegraphics[width=200pt,angle=0.]{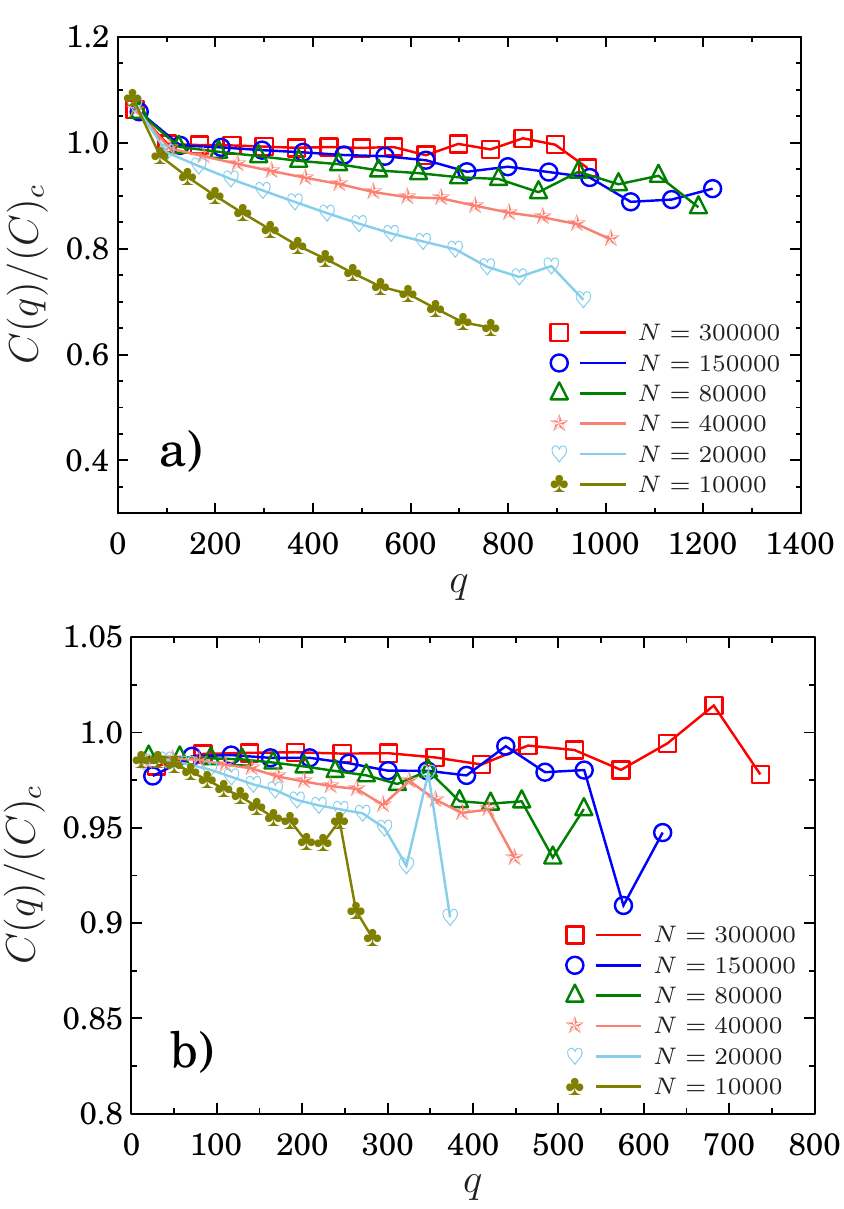}
\caption{(Color online) Degree dependence of the local clustering coefficient of sparse equilibrium networks of mean degree $\langle q \rangle = 20$
and different sizes. (a) model 1, (b) model 2. The local clustering coefficient is normalized by its value for the corresponding uncorrelated network.}
\label{fig:sparse_C}
\end{figure}

To study correlations and clustering, we measured the degree dependence of the average degree of nearest neighbours and the clustering
coefficient (Figs. \ref{fig:sparse_qnn} and \ref{fig:sparse_C}). In the plots we normalized the measured values
$\overline{q}_{nn}(q)$ and $C(q)$ by the values expected in the uncorrelated case. These corresponding uncorrelated
values, denoted by $(\overline{q}_{nn})_c$ and $(C)_c$, are just the values calculated in the configuration model
using the same structural characteristics 
$N$, $\langle q \rangle$, and $\langle q^2 \rangle$ as those obtained in the simulations:
\begin{eqnarray}  
(\overline{q}_{nn})_c &=& \frac{\langle q^2 \rangle}{\langle q \rangle} 
,
\label{eq:qnn} 
\\[5pt]
  (C)_c &=& \frac{1}{N \langle q \rangle} \left\lgroup \frac{\langle q^2 \rangle - \langle q \rangle}{\langle q \rangle} \right\rgroup ^2 
  .
  \label{eq:clus}
  \end{eqnarray}


Figures \ref{fig:sparse_qnn} and \ref{fig:sparse_C} 
confirm a convergence to an uncorrelated equilibrium state for large
networks. It is interesting to note that although correlations are smaller in the second
model, Fig. \ref{fig:sparse_qnn}(b), compared to the first one, Fig. \ref{fig:sparse_qnn}(a), the degree distributions in
the second model at the same sizes are still further away from the form of Eq. (\ref{eq:solution}). 
Model 1 exhibits stronger correlations and stronger degree dependence of the local clustering coefficient (Fig. \ref{fig:sparse_C}(a) compared with
Fig. \ref{fig:sparse_C}(b)), even though
the degree distributions in model 1 (Fig. \ref{fig:sparse_dd}(a)), for the system sizes considered, already practically coincide with the uncorrelated form of Eq. (\ref{eq:solution}).


\subsection{Denser networks} 

\begin{figure}[t]
\centering
\includegraphics[width=200pt,angle=0.]{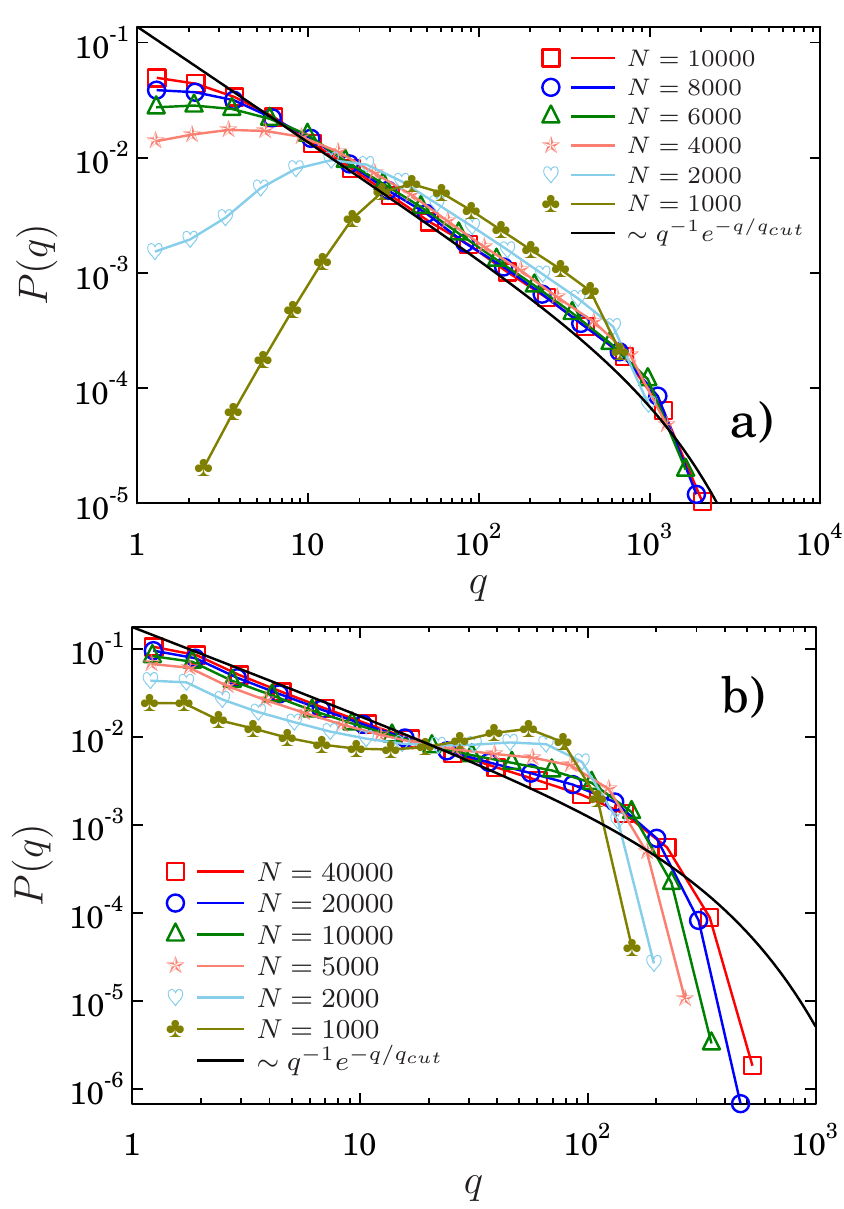}
\caption{(Color online) Degree distribution of denser (than in Fig.~\protect\ref{fig:sparse_dd}) equilibrium networks of different sizes.
(a): $\langle q \rangle = 200$, model 1; (b): $\langle q \rangle = 50$, model 2.}
\label{fig:dense_dd}
\end{figure}

\begin{figure}[t]
\centering
\includegraphics[width=200pt,angle=0.]{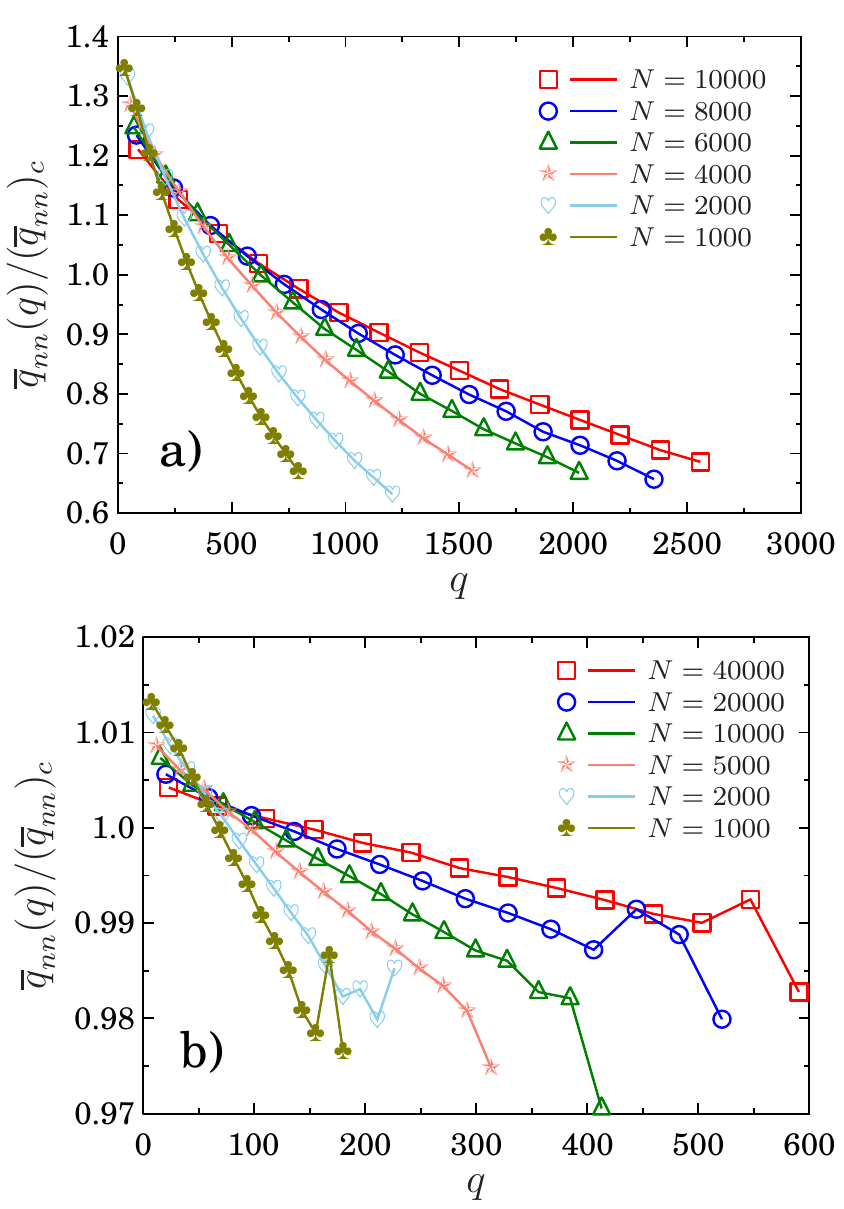}
\caption{(Color online) Degree dependence of the mean degree of nearest neighbours of dense equilibrium networks of varying size.
(a): $\langle q \rangle = 200$, model 1; (b): $\langle q \rangle = 50$, model 2.}
\label{fig:dense_qnn}
\end{figure}

\begin{figure}[!]
\centering
\includegraphics[width=200pt,angle=0.]{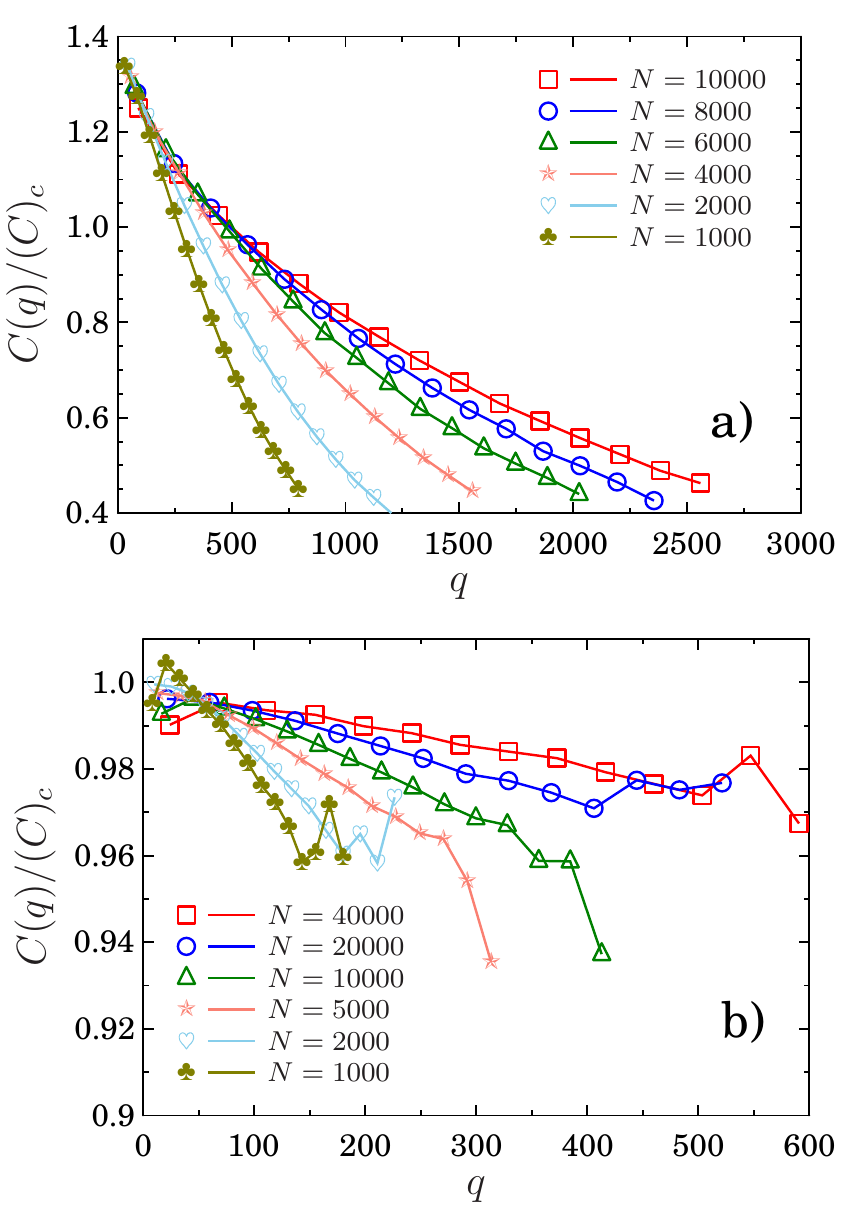}
\caption{(Color online) Degree dependence of the clustering coefficient of dense equilibrium networks of varying size.
(a): $\langle q \rangle = 200$, model 1; (b): $\langle q \rangle = 50$, model 2.}
\label{fig:dense_C}
\end{figure}

We performed simulations of networks with higher mean node degrees, $200$ and $50$ (models 1 and 2, respectively), than in the preceeding subsection. This enabled us to observe stronger size effects in the degree distributions, Fig.~\ref{fig:dense_dd}(a), (b), than in Fig.~\ref{fig:sparse_dd} at the same network sizes. Simulations for model 2 are particularly time consuming, so the mean degree, $50$, has to be chosen smaller than $200$ for model 1.
The system sizes were chosen in a way to capture a wide range of behaviors in both models, using the highest possible mean degree (limited by computational time).
Figures~\ref{fig:dense_dd}(a), (b) demonstrate markedly different degree distributions for models $1$ and $2$ at low network sizes. Note that the difference is not only in a hump present in Fig.~\ref{fig:dense_dd}~(b) but this difference is well observable even in the range of small degrees. The degree-degree correlations for these networks demonstrate a stronger disassortative mixing, Fig.~\ref{fig:dense_qnn}(a), (b), than for their less dense counterparts in Fig.~\ref{fig:sparse_qnn}(a), (b). 
The degree dependence of the local clustering coefficient is also more pronounced in denser networks, Fig.~\ref{fig:dense_C}(a), (b), than in their less dense counterparts, Fig.~\ref{fig:sparse_C}(a), (b). This is especially well seen for model~1, compare respective Fig.~\ref{fig:dense_C}(a) ($\langle q \rangle=200$) and Fig.~\ref{fig:sparse_C}(a) ($\langle q \rangle=20$).



\subsection{A real-world example}

\begin{figure}[t]
\centering
\includegraphics[width=200pt,angle=0.]{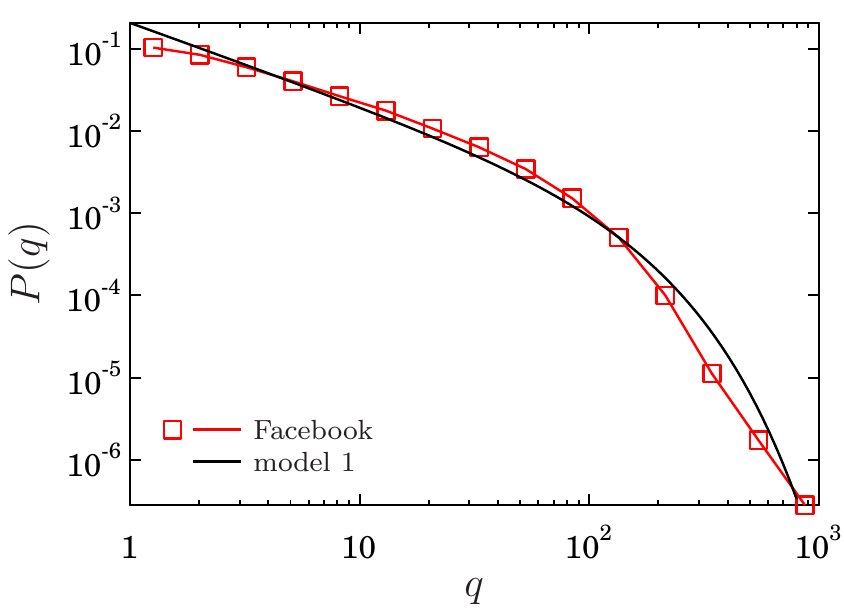}
\caption{(Color online) Degree distribution taken from a Facebook sample, compared to the degree distribution of model 1, using the
same system size and mean degree: $N = 63731$ and $\langle q \rangle = 25.64$.}
\label{fig:facebook}
\end{figure}

To demonstrate that degree distributions of $\gamma = 1$ do exist in reality, we 
explore data from Facebook. The analyzed sample
consists of all of the user-to-user links from the Facebook New Orleans (2009) networks \cite{viswanath-2009-activity}.
This sample has size $N = 63731$ and mean degree $\langle q \rangle = 25.64$. Figure~\ref{fig:facebook} shows the measured
Facebook degree distribution and the degree distribution from our model 1 using the same system size and mean degree.
Notice the closeness of the two curves although no fitting was done. 
In this parameter range, the degree distribution of the model is already very close to the analytical form given by
Eq. (\ref{eq:solution}).

We see that the curve from our model 1 provides a good approximation of the empirical distribution.
We stress 
that the underlying structures of the two networks are different, and so our models cannot be applied directly. 
Facebook is growing, and like the majority of social networks, it exhibits assortative correlations, while our equilibrium models produce disassortative ones.
The Pearson correlation coefficients are $0.175$ and $-0.004$ for the Facebook sample and our model 1, respectively; with these parameters model 1 is already close to the uncorrelated sparse limit.
Also, social networks have strong clustering, whereas our models have very low clustering coefficients for large
systems (and clustering actually vanishes in the infinite size limit).
The corresponding clustering coefficients are $0.148$ and $0.006$.
Nevertheless, Fig.~\ref{fig:facebook} indicates that such low exponents of the degree distribution do appear in reality. 
Therefore it may be 
useful to think outside the realm of conventional preferential attachment models in order to come
closer to a full explanation of real-world network structures.


\section{Discussion} 

Previously studied network models generating degree distributions with exponents $\gamma$ smaller than $2$ exploited a set of rather intricate mechanisms
and non-trivial ideas. In particular, these included fitness models \cite{bianconi2001competition,bianconi2001bose}, accelerated growth, where the network
becomes denser with time \cite{dorogovtsev2002evolution,dorogovtsev2003evolution}, aggregation processes \cite{seyed2006scale}, the power of
choice \cite{d2007power}, etc. At first sight, the two equilibrium network models that we have considered in this paper are simpler. In the infinite size limit,
both these rewiring models generate uncorrelated networks with the degree distributions
$P(q) \sim q^{-1} e^{-q/(\langle q \rangle \ln\langle q \rangle)}$. In finite networks, however, these models become essentially non-trivial due to constraints
for rewiring which occur in this situation. We have found that these constraints lead to strong disassortative degree--degree correlations and to
degree-dependent local clustering. They also markedly change the form of the degree distributions. The structural constraint is particularly strong for model 2,
so the results for these two closely related models in finite systems significantly differ from each other. 

Finally, we emphasize a strong difference of these rewiring models from well studied equilibrium networks based on the preferential attachment
mechanism \cite{dorogovtsev2003principles}. While networks in the present work demonstrate a  power-law degree distribution in a wide range of mean degrees,
the networks from Ref.~\cite{dorogovtsev2003principles} are scale-free only at a critical mean degree value.

The resulting degree distributions are observable only if the mean number of connections of nodes in a network is sufficiently large.
This is the case for a number of real-world networks, including social and neural networks.
(The mean number of friends of adult Facebook users was already 338 in 2014 \cite{facebook2014} and the mean number of synapses in brain
neuronal networks is generally of the order of $10^3$ \cite{herculano2006cellular, herculano2012remarkable}).
We suggest that our results may be useful for understanding the structural properties of networks of this kind.

\section*{Acknowledgement}

This work was supported by the FET proactive IP project MULTIPLEX 317532.

\section*{References}

%

\end{document}